# Transformative Effects of ChatGPT on Modern Education: Emerging Era of AI Chatbots

Sukhpal Singh Gill[1], Minxian Xu[2], Panos Patros[3], Huaming Wu[4], Rupinder Kaur[5], Kamalpreet Kaur[6], Stephanie Fuller[7], Manmeet Singh[8,9], Priyansh Arora[10], Ajith Kumar Parlikad[11], Vlado Stankovski[12], Ajith Abraham[13], Soumya K. Ghosh[14], Hanan Lutfiyya[15], Salil S. Kanhere[16], Rami Bahsoon[17], Omer Rana[18], Schahram Dustdar[19], Rizos Sakellariou[20], Steve Uhlig[1] and Rajkumar Buyya[21]

[1]School of Electronic Engineering and Computer Science, Queen Mary University of London, London, UK
[2]Shenzhen Institute of Advanced Technology, Chinese Academy of Sciences, Shenzhen, China
[3]Raygun Performance Monitoring, Wellington, New Zealand
[4]Center for Applied Mathematics, Tianjin University, Tianjin, China
[5]Department of Science, Kings Education, London, UK
[6]Cymax Group Technologies, British Columbia, Canada
[7]QM Academy, Queen Mary University of London, London, UK
[8]Jackson School of Geosciences, University of Texas at Austin, Austin, TX, USA
[9]Centre for Climate Change Research, Indian Institute of Tropical Meteorology, Pune, India
[10]Microsoft, Hyderabad, India
[11]Institute for Manufacturing, Department of Engineering, University of Cambridge, Cambridge, UK
[12]Faculty of Computer and Information Science, University of Ljubljana, Ljubljana, Slovenia
[13]Machine Intelligence Research Labs, Auburn, WA, USA
[14]Department of Computer Science and Engineering, Indian Institute of Technology, Kharagpur, India
[15]Department of Computer Science, University of Western Ontario, London, Canada
[16]School of Computer Science and Engineering, The University of New South Wales (UNSW), Sydney, Australia
[17]School of Computer Science, University of Birmingham, UK
[18]School of Computer Science and Informatics, Cardiff University, Cardiff, UK
[19]Distributed Systems Group, Vienna University of Technology, Vienna, Austria
[20]Department of Computer Science, University of Manchester, Oxford Road, Manchester, UK
[21]Cloud Computing and Distributed Systems (CLOUDS) Laboratory, School of Computing and Information Systems, The University of Melbourne, Australia

*s.s.gill@qmul.ac.uk, mx.xu@siat.ac.cn, patrospanos@gmail.com, whming@tju.edu.cn, rupinderchem@gmail.com, kamalpreet.k@cymax.com, stephanie.fuller@qmul.ac.uk, manmeet.singh@utexas.edu, priyansh.arora@microsoft.com, aknp2@cam.ac.uk, vlado.stankovski@fri.uni-lj.si, ajith.abraham@ieee.org,, skg@cse.iitkgp.ac.in, hanan@csd.uwo.ca, salil.kanhere@unsw.edu.au, r.bahsoon@cs.bham.ac.uk, ranaof@cardiff.ac.uk, dustdar@dsg.tuwien.ac.at, rizos@manchester.ac.uk, steve.uhlig@qmul.ac.uk, rbuyya@unimelb.edu.au*

**Abstract**

ChatGPT, an AI-based chatbot, was released to provide coherent and useful replies based on analysis of large volumes of data. In this article, leading scientists, researchers and engineers discuss the transformative effects of ChatGPT on modern education. This research seeks to improve our knowledge of ChatGPT capabilities and its use in the education sector, identifying potential concerns and challenges. Our preliminary evaluation concludes that ChatGPT performed differently in each subject area including finance, coding and maths. While ChatGPT has the ability to help educators by creating instructional content, offering suggestions and acting as an online educator to learners by answering questions and promoting group work, there are clear drawbacks in its use, such as the possibility of producing inaccurate or false data and circumventing duplicate content (plagiarism) detectors where originality is essential. The often reported "hallucinations" within Generative AI in general, and also relevant for ChatGPT, can render its use of limited benefit where accuracy is essential. What ChatGPT lacks is a stochastic measure to help provide sincere and sensitive communication with its users. Academic regulations and evaluation practices used in educational institutions need to be updated, should ChatGPT be used as a tool in education. To address the transformative effects of ChatGPT on the learning environment, educating teachers and students alike about its capabilities and limitations will be crucial.

## 1. Emerging Era of AI Chatbots

Artificial Intelligence (AI) is becoming increasingly prevalent in various sectors, including higher education. AI applications are becoming crucial for colleges and universities, whether it be for personalised learning, computerised assessment, smart educational systems, or supporting teaching staff. They offer support that results in reduced expenses and enhanced learning results. Chatbots are AI-powered software applications that can mimic human conversational interactions [1]. They function by assessing a conversation's context and coming up with answers they think are appropriate. They can answer a variety of problems since they have been taught using large linguistic datasets [2]. A wide range of educational organisations, from elementary and secondary schools to universities and professional development programmes, can benefit from using chatbots such as ChatGPT and Google Bard [3]. Their capacity to provide personalised education is one of their strongest points.

This article addresses ChatGPT, an AI assistant chatbot [1], released by OpenAI [2], that has received significant public attention since its introduction in November 2022 – amongst professionals, students, policymakers, and experts in the field of higher education. With both hope and prudence, there is still an open debate on the role of AI technologies and their appropriateness in education, and their influence on learning, students' development, evaluation and assessment, and certification, especially in human-led teaching. Whilst some educators and practitioners see opportunities that tools such as ChatGPT can facilitate learning and development and have been calling for regulation, others see threats for the core mission of education: development and training to acquire knowledge and problem solving skills that could be applicable in a wide range of domains and time varying contexts, fairness in assessment, certification and meaningfulness of the education awards, and making inequality in education worse [3] Despite the divide, ChatGPT has seen the greatest consumer growth since its introduction and currently supports more than 100 million active users [4].

ChatGPT uses AI and Natural Language Processing (NLP) to respond to user input queries and generate answers that are *human-like*. It has drawn international interest because of its efficacy in generating answers that are cogent, orderly and instructive. Notwithstanding its popularity, ChatGPT has created fresh difficulties and risks for education. There are concerns regarding the potential misuse of AI Generated Content (AIGC), as it could be employed to generate academic tests and assignments for students and provide tailored responses to coursework questions and assessments. As a result, a number of institutions have forbidden students from using ChatGPT [27] – including a ban within an entire country. Researchers [1, 3, 4, 5] have examined the effects of ChatGPT on learning and found that teachers were worried about using it in the classroom. They voiced concerns that since ChatGPT can quickly produce appropriate content, learners may utilise it for outsourcing their assignments. Further, a number of issues have been identified, including copied content, wrong replies and improper referencing (or no referencing at all). Hence, it is crucial to carefully examine the impact of ChatGPT-assisted education to fully leverage its benefits while mitigating any drawbacks. This is not a new phenomenon, as the launch of search engines led to similar types of concerns. However, with a search engine, a content user can cross reference specific URLs that have been used to achieve a particular outcome. In ChatGPT no specific references or URLs are included in the generated text. Figure 1 shows the benefits of ChatGPT to a wide range of companies, software developers, and end users within the education sector.

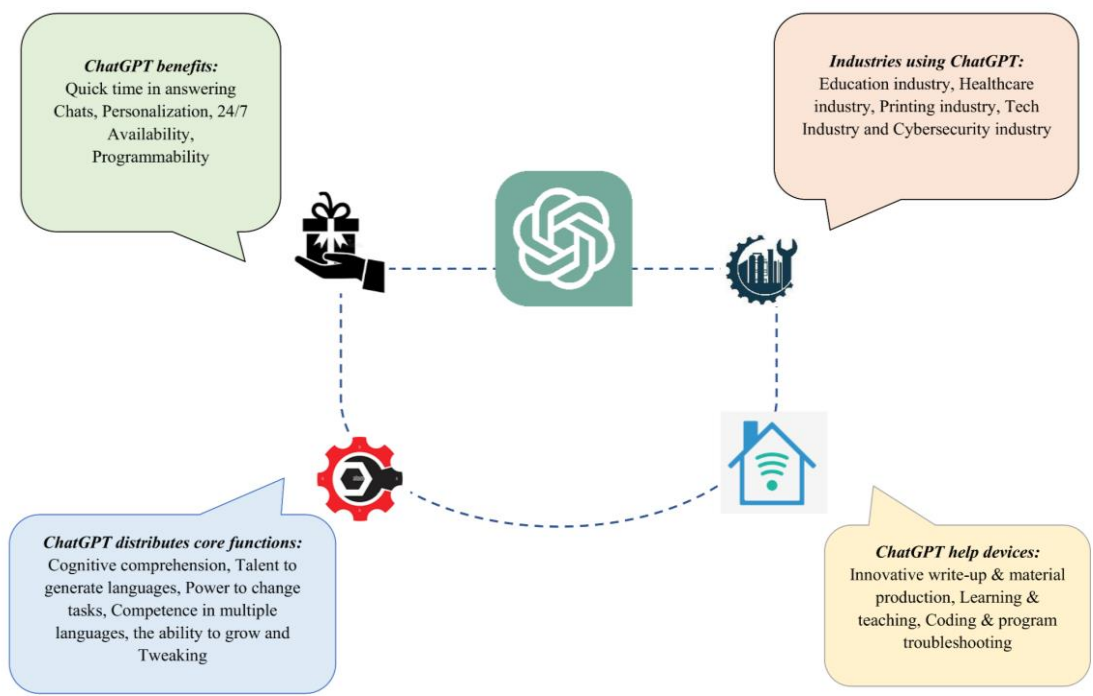

Figure 1: ChatGPT benefits a wide range of companies, software developers and end users within the education sector

## 2. Transforming Online Education: The Role and Impact of ChatGPT

In the evolving landscape of online education, artificial intelligence (AI) is proving to be a game-changer, with tools such as ChatGPT leading a radical transformation that ranges from assessment design to language learning. Many e-learning platforms, including Coursera, use AI to detect common errors in their assignments [6]. Using ChatGPT to produce material, instructors can develop unique assessments and learning content, while businesses (e.g., Course Hero) could see their paid homework assistance models disrupted by the AI's cost-effective versatility [7]. AI chatbots can assist students in honing their understanding of text by posing personalised queries and offering comments on their responses. The tool can also be used to enhance a person's critical and analytical abilities. Another crucial aspect of ChatGPT is helping learners explore languages, enabling students to modify sentences, practise correct pronunciation and terminologies, grasp sentence structure and give real-time interpretations [8]. Perhaps the most significant and controversial application of ChatGPT is to produce written content in response to exam or essay questions. It enables both educators and students to to compose articles on any subject in response to prompts that are input to this software. Additionally ChatGPT can offer suggestions for how to enhance grammatical structures, brevity, or clearness of multiple drafts of the same essay – enabling users to get through barriers to writing and provide fresh viewpoints on their selected subject.

The outcomes for both ethics and business are huge. Content authors who use ChatGPT may reconsider paying steep membership fees for AI creative solutions like Grammarly. According to the Gartner Hype Cycle, ChatGPT is presently at the "trough of disillusionment" stage, in which individuals lack hope. After research and refining, the technology enters the "plateau of productivity"[1], when customers and organisations recognise, welcome, and utilise it. [10]. Nevertheless, OpenAI hopes to improve effectiveness, expand its range of skills, and commercialise new advancements through its cooperation with Microsoft [2]. The release of ChatGPT4 in March 2023 has already generated debate since the tool looks more precise, dependable, and responsive [11]. Adoption will rise as a result of these upgrades and better interaction with different productivity applications such as Slack, or Zoom [9]. As the AI-powered educational tools continue to advance, embracing and understanding their use becomes crucial; the advent of ChatGPT4 highlights this, setting the stage for more precise, reliable, and responsive education technology.

---
[1] https://analyticsindiamag.com/how-chatgpt-broke-the-ai-hype-cycle/

## 3. Educating with ChatGPT

As an initial foundation for developing course syllabi, instructional resources, and evaluation activities, ChatGPT might be a useful tool for educators. However, there are issues that need to be resolved about the produced content's authenticity [12]. A potential fix might be to build training materials for course-specific bots using ChatGPT. For instance, utilising ChatGPT to support students' acquisition of the English language via taking the role of a "native-speaker" the student can converse with [13]. After making sure the materials are accurate, teachers may ask ChatGPT to transform them into a layout compatible with AI-based chatbots, giving students a customised and engaging learning experience. Additionally, ChatGPT can improve techniques for active learning. By way of illustration, flipped education can be employed, in which learners are expected to read material before classes. This kind of education can allow for more participatory learning activities, including group discussions during class time. Nevertheless, in traditional flipped classrooms, students can struggle with pre-class learning [14]. This problem was made clear during the COVID-19 pandemic [26], when entirely online education resulted in subpar participation in the classroom and disinterest in peer sharing [7, 9, 15]. ChatGPT, as a virtual instructor, may help learners with their web-based independent research by responding to their inquiries and can improve collaboration by offering suggestions for a debate framework and giving immediate response.

## 4. Higher Education Risks from ChatGPT

Researchers claim that there are issues with ChatGPT's reliability and precision that make it difficult to employ in educational settings. ChatGPT could be biased or inaccurate because it was trained on such a big amount of data [16]. Further, bias may result from using studies that were predominantly done in nations with high income or controversial books that did not appeal to everyone. Moreover, ChatGPT has little information and has not (yet) been fully upgraded with information after 2021 [16, 17]. Thus, especially for specific topics and current events, its comments could occasionally not be precise or dependable. Additionally, ChatGPT may produce inaccurate or incorrect data. For learners who depend on ChatGPT, any inaccuracies would not only disrupt the learning process but could also jeopardise the integrity and credibility of the educational experience, thus breaking the trust-bond that is fundamental to effective education.

The issue of AI-generated content being passed for original student work has grown significantly. Investigations show that ChatGPT can get past the usual plagiarism detection tools such as Turnitin by producing information that appears to be unique. Students who utilised ChatGPT were more likely to plagiarise than those who did not, according to the literature [18, 19]. This is a serious challenge to academic credibility and the valid and fair assessment of student learning [20]. Even where use of ChatGPT is permitted within assessment, learners who utilise it have an unfair edge over other learners who have no opportunity to do so. More significantly, while using ChatGPT, teachers may be less able to assess how students perform with accuracy, which makes it challenging to monitor students' learning issues.

It is crucial to recognise the risks chatbots may bring in the context of digital destitution and the technological divide, despite the fact that they have the ability to improve education [28]. Chatbots and other technological resources for learning could not be available for learners who do not have the ability to utilise a stable Internet connection or do not have the materials necessary to engage in virtual classes. To combat this, learning institutions need to take preventative steps to guarantee that all learners have equitable access to digital tools like chatbots [29]. In addition, it might be difficult to get learners to see that every individual has the same opportunity to utilise technology which they have. Teachers may play an important role in spreading this comprehension by including discussions of digital inequality and the need for universal access in their curricula [30]. Finally, colleges and universities can partner with local groups that help learners who need assistance by providing them with Internet connectivity or laptops to use for free.

## 5. Treatment Required Immediately in Reaction to the Effects of ChatGPT

The issues brought about by the advent of artificially intelligent material in educational assignments call for the updating of assessment practices and institutional norms. Teachers should improve the structure of the exams/assignments by utilising interactive tools to lower the possibility of copying. The inability of ChatGPT's initial version to handle photos and videos led to the absence of context, which made it more difficult for learners who wanted to make use of it for illicit purposes. Nevertheless, OpenAI's GPT-4, an enormous multimodal model, can analyse pictures [2, 11]. Learners must be present physically (face to face interaction) and in fact show their skills for these elements. AI-based writing identification technologies could be made accessible to educators at the organisational level. To define the limits of ChatGPT's engagement in students' learning, combating plagiarism rules must be defined.

Reacting to ChatGPT's effects also requires teacher development and education for learners [21]. It is crucial to advise professors on how to design out opportunities for plagiarism in assessment. Training can also be provided to spot ChatGPT use in learner tasks by utilising AI content detection toolkits. Teachers should also receive training in order to properly utilise ChatGPT in their lesson preparation and programme evaluation. It is vital to educate learners about ChatGPT's drawbacks, including its reliance on skewed data, lack of current knowledge, and possibility for producing false or misleading findings [22]. The factual veracity of the material offered by ChatGPT should thus be checked, evaluated, and corroborated using other reliable sources, such as reference books, according to teachers [23]. Increasing students' knowledge of academic honesty guidelines and their comprehension of the penalties for misconduct in academia are also significant [24]. Educators should publicly address ChatGPT in their classes and stress the value of academic integrity in order to accomplish this aim [25].

## 6. Future Perspectives and Potential Challenges

In what follows, there are some key challenges for the foreseeable future that we have identified based on our comprehensive understanding of AI and expertise in the education sector:

- Limiting or forbidding ChatGPT won't solve the problem. It is preferable to accept it and establish explicit guidelines for its application in academic settings [4,11,19].
- The danger posed by ChatGPT to traditional evaluation methods is substantial [8, 12, 14]. Future evaluation models must, without a doubt, place a greater emphasis on autonomous and reflective thought, logical deduction, complex problem-solving that expects the use of AI tools, innovative thinking, and challenging and confirming data inputs.
- More monitoring and openness will be required as ChatGPT develops in order to distinguish between information produced by humans and AI [9, 17]. Regulators are now far behind technical advancement, but the discussion sparked by all parties involved will undoubtedly lead to legislative action.
- In the context of prompt engineering, innovative methods are required to refine and optimise the queries posed to ChatGPT to ensure that the responses are accurate, engaging and relevant.
- Students with special needs may encounter additional barriers while attempting to use a chatbot [32]. Learners with intellectual disabilities might require extra help understanding and using chatbots, whereas learners with visual impairments may have trouble seeing chatbot content. Instructors need to make sure learners who struggle can get the assistance they need.
- The discussion expands far beyond the limited scope of the education sector: in today's business world, learning and assessment are ongoing processes. Parallel opportunities and challenges arise, such as when a software engineer is assessed and hired through a coding challenge or when a new hire is integrated into an organisation's business and technical operations. If ChatGPT can assist developers in creating more stable software in a shorter amount of time, it might be of great benefit. However, it is our responsibility as educators to teach the software engineer how to manage and assess technology for optimal results, which necessitates a transparent trail of actions.
- One potential route to enhance Large Language Models (LLM) could be by using stochastic models, such as Markov Decision Process (MDP) in their design. This is a subject of current research in the context of the ExtremeXP project.[2] This research aims at building more sincere and sensitive AI-based systems.
- To finish on an optimistic note, we think that the process of learning will concentrate more on the human components of the method by depending more on ChatGPT or similar AI technologies in future in order to automate various tasks such as study instructions, descriptions, diagrams of concepts, practise exams and quizzes [12]. The ultimate objective should be to shift away from memorising and towards developing the durable skills that a futuristic humanity would require. As identified in the literature [9, 13, 16], using ChatGPT properly will enable teachers and students to employ more comprehensive teaching and assessment strategies, like project-oriented, inquiry-based, cooperative, personalised, or multidisciplinary education.

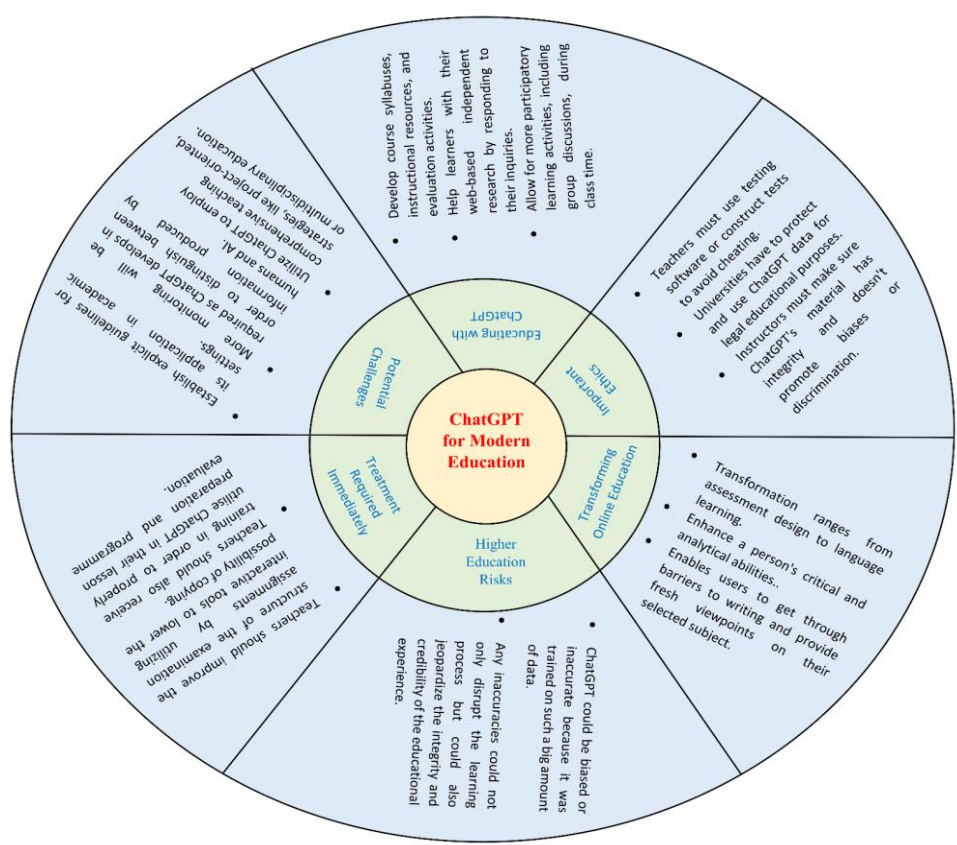

Figure 2: Transformative effects of ChatGPT on modern education

---

[2] https://extremexp.eu/

## 7. Summary of Findings and Take-Aways

This article discussed the transformative effects of ChatGPT on modern education. Figure 2 provides a brief outline of the key results and takeaways that educators can utilise to ensure the effective implementation of Chatbots like ChatGPT in the educational settings. This discussion brought to light ChatGPT's lack of consistency across many topic areas and its possible advantages when acting as an Internet-based tutor to learners and an assistance for lecturers. Nonetheless, its usage creates a number of issues, including the production of false or inaccurate material and the danger it brings to academic credibility. Since ChatGPT was trained on quite a massive quantity of data, it raises serious ethical concerns that it may be prejudiced or erroneous; this might not only compromise the quality of the education, but it could additionally cast doubt on the reliability of the data used to train the system. The results of this research demand that schools and institutions change their standards and procedures for preventing plagiarism right now, while moving to update their learning and evaluation methods to include rather than being apprehensive to AI. Teachers should receive training on how to utilise ChatGPT efficiently and spot plagiarism in homework. Learners must also be made aware of ChatGPT's capabilities, restrictions, and possible impact on their educational credibility. In this respect, calls to regulate ChatGPT's use need to be considered seriously not only to mitigate the risks but also to help all stakeholders prepare as necessary.

**Declaration of competing interest**

The authors declare that they have no known competing financial interests or personal relationships that could have appeared to influence the work reported in this paper.